\documentclass[%
 aapm,
 amsmath,amssymb,
 reprint,%
]{revtex4-2}

\usepackage{graphicx}
\usepackage{dcolumn}
\usepackage{bm}

\usepackage[mathlines]{lineno}
\modulolinenumbers[5]

\begin{document}

\preprint{AAPM/123-QED}

\title[]{Effects and side-effects of plasmonic photothermal therapy in brain tissue}

\author{Yue He, Kristoffer Laugesen, Dana Kamp, Salik Ahmad Sultan, Lene Broeng Oddershede and Liselotte Jauffred}
 \email{jauffred@nbi.dk}
\affiliation{The Niels Bohr Institute, University of Copenhagen, Copenhagen, DK-2100, Denmark}

\date{\today}

\begin{abstract}
Heat generated from plasmonic nanoparticles can be utilized in plasmonic photothermal therapy. A combination of near-infrared laser and metallic nanoparticles are compelling for the treatment of brain cancer, due to their efficient light-to-heat conversion and bio-compatibility. However, one of the challenges of plasmonic photothermal therapy is to minimize the damage of the surrounding brain tissue. The adjacent tissue can be damaged as the results of either absorption of laser light, thermal conductivity, nanoparticles diffusing from the tumor, or a combination hereof. Hence, we still lack the full understanding about the light-tissue interaction and, in particular, the thermal response. We tested the temperature change in three different porcine cerebral tissues, i.e., the stem, the cerebrum, and the cerebellum, under laser treatment. We find that the different tissues have differential optical and thermal properties and confirm the enhancement of heating from adding plasmonic nanoparticles. Furthermore, we measure the loss of laser intensity through the different cerebral tissues and stress the importance of correct analysis of the local environment of a brain tumor.
\end{abstract}

\maketitle
\section*{Introduction}
Brain malignant tumors are either primary cancers of the central nervous system or secondary tumors arising from metastasis \cite{Louis2016}. Conventional treatments include surgery, radiotherapy, chemotherapy, and symptomatic therapies. Despite this, brain tumors are still associated with poor prognosis, for instance due to recurrence from the resection site. So because of the infiltrative growth of the tumor in the surrounding (and healthy) brain tissue, any potential therapy must take care not to damage healthy tissue \cite{Sedo2012}. 

Recent advances in nanotechnology bears promise of using nanoparticles in cancer therapy as light-to-heat converters for plasmonic photothermal therapy (PPTT) \cite{Huang2008}. The therapy is based on the following principle: The electromagnetic field of a laser beam resonates with the plasmonic field of nanoparticles, which absorb the incident light. The absorbed energy is dissipated as heat within a distance comparable to the particle’s diameter \cite{Bendix2010}.
Hence, if nanoparticles are embedded in tumors, laser irradiation will cause hyperthermia (41$^{\circ}$-44$^{\circ}$ C) and, thus, localized irreversible damage of the cancerous tissue \cite{ONeal2004}. This effect is further enhanced by the fact that cancer cells are more susceptible to thermal treatment than healthy cells \cite{Zee2002}. Our choice of plasmonic nanoparticles is gold nanoshells (AuNSs) with a core diameter of 120 nm and 15 nm shell that are resonant with near-infrared (NIR) lasers (810 nm) in the the biological transparency window \cite{Jaque2014}. Furthermore, AuNSs have been shown promising for PTTP both \textit{in-vitro} \cite{Qin2012} and \textit{in-vivo} \cite{Huang2010,Jorgensen2016,Yang2016}.
\begin{figure*}
\centering
\includegraphics[]{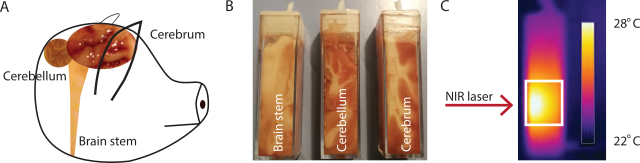}
\caption{\label{fig:1}\textbf{Experimental setup}. 
(\textbf{A}) Sketch of a pig brain's cross section illustrating the position of the brain stem, cerebrum, and cerebellum. The structures shown here have symmetric counterparts in the other half of the brain. 
(\textbf{B}) Cuvettes with dissected cerebral tissue: Brain stem, cerebrum, and cerebellum, respectively. 
(\textbf{C}) Thermal image of a cuvette containing cerebral tissue irradiated with a NIR laser (red arrow) for 25 minutes. The region of interest (white box) encloses the pixels analyzed to find the maximum temperature which is denoted $T$. The color bar denotes absolute temperatures.}      
\end{figure*}
%

One challenge, however, of PPTT is the particle delivery to the tumor. In general there are two strategies either via inter-tumoral or intravenous injections. 
To improve circulation and thereby increase accumulation, it has been suggested to coat the nanoparticles with a shielding bio-membrane \cite{Rengan2015,Xuan2016} or therapeutic drugs that can be released upon laser irradiation \cite{Kennedy2011a,Huang2014,Zhou2015}. 
An important aspect of PTTP that has attracted less attention, is the inevitable radiation of the healthy tissue surrounding the tumor and the resulting laser induced damage of healthy tissue. 
This interaction between light and tissue depends on the scattering and absorption properties of tissue as well as on the  thermal conductivity and diffusivity of cerebral tissue \cite{Jacques2013,Yaroslavsky2002}. 
 
In this study, we measure the temperature change in a phantom mimicking brain tissue and also in different porcine cerebral tissues. For all samples types investigated, we measure the temperature under laser irradiation and we find a significant difference between the heating of brain stem and the other types of tissue upon laser irradiation. Also, we find a increased heating in tissue injected with AuNSs upon laser irradiation, hence, the positive treatment outcome is affirmed, however the effect is diminished by the strong absorption and scattering of the the brain tissue itself. 
Thus, we conclude that the effect of PPTT on the tumor and the side-effects on the healthy brain tissue is related in a non-trivial manner; depending on the tissue properties.

\section*{Results}
To measure the light to heat conversion we prepared two different kinds of samples; one artificial brain phantom and one from dissected pig brain (Fig. \ref{fig:1}(A)). The samples were irradiated with a NIR laser with a wavelength of 806 nm and the temperature (Fig. \ref{fig:1}(B)) was recorded by an infrared (thermal) camera. Fig. \ref{fig:1}(C) is an example of such a thermal image, where the temperature, $T$, is defined as the maximum temperature within the region of interest (white square).

\subsection*{Plasmonic photothermal heating of artificial tissue}
\begin{figure}
\includegraphics[]{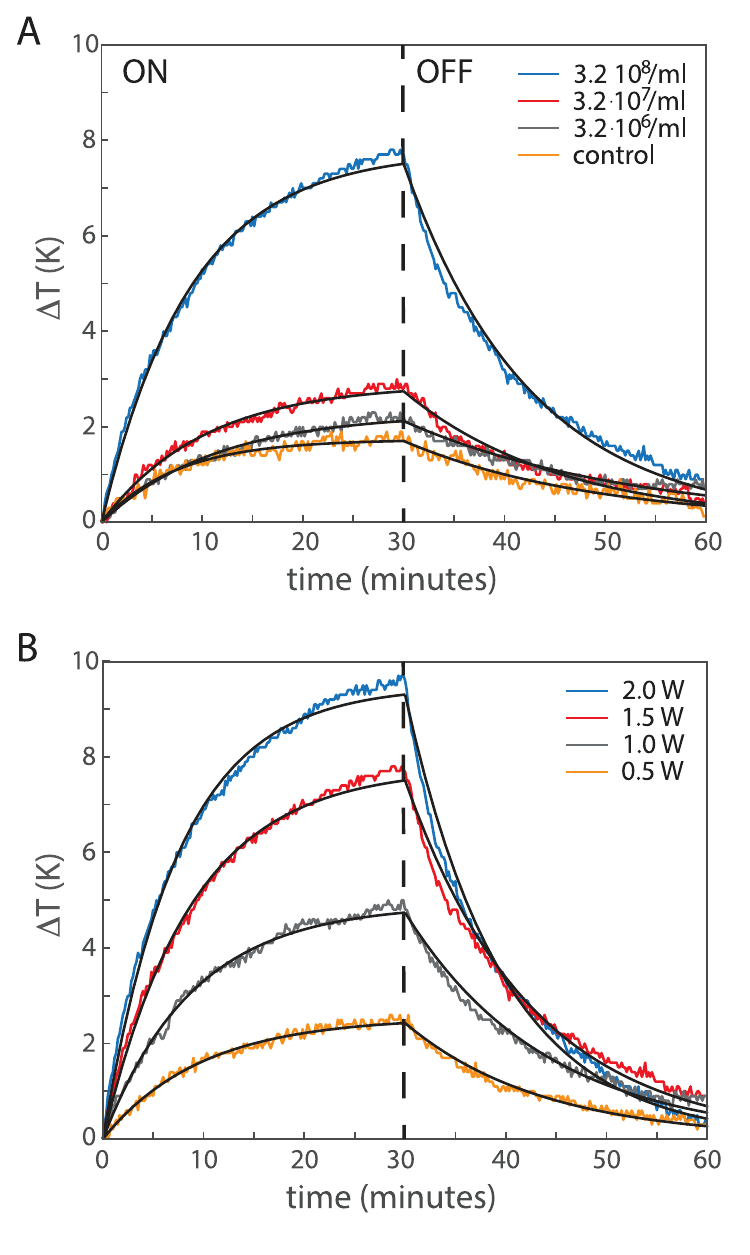}
\caption{\label{fig:2}\textbf{Plasmonic photothermal heating of artificial tissue}.
Plots of temperature versus time for artificial tissue using different AuNS concentrations and laser powers. The  806 nm laser irradiation (ON) and without (OFF). The vertical black punctuated line signifies the instance where the laser is turned off and the full black lines are fits of Eq. \eqref{heatFit} and Eq. \eqref{coolFit} to data.
(\textbf{A}) Temperature change, $\Delta T$, versus time for different AuNSs concentrations (see legends) and a control with no AuNSs (yellow). The laser power, $P$, is 1.5 W. 
(\textbf{B}) Temperature change versus time for 5 different laser powers (see legends) and a AuNSs concentration of $3.2\times10^8$ particles/ml.}
\end{figure}

We experimentally determined the heating of different concentrations of AuNSs in artificial tissue, i.e., a 0.6\% agarose solution \cite{Eldridge2016}, and a control without AuNSs. Fig. \ref{fig:2}(A) shows temperature change, $\Delta T$, versus time, $t$, of the phantom tissue for varying particle concentration. The tissue was heated with a 1.5 W laser for half an hour and cooled (laser off) for another half an hour.
The temperature driving force approximation \cite{Bardhan2009,Roper2007,Pattani2013} permits the system equilibrium time, $\tau$, and the steady-state temperature, $T_{ss}$, to be estimated using system parameters from heating data and cooling data, respectively. For the heating we found the following exponential dependence on $t$: 
\begin{equation}\label{heatFit}
\Delta T=\Delta T_{ss}(1-e^{-t/\tau})    
\end{equation}
and for cooling:
\begin{equation}\label{coolFit}
\Delta T=\Delta T_{ss}e^{-t/\tau}.    
\end{equation}
It is worth noticing that Eq. \eqref{heatFit} and Eq. \eqref{coolFit} both are independent of the ambient temperatures. Thus, the results from this study, which were all obtained at room temperature, are also applicable at body temperature (37$^{\circ}$ C). For the phantom tissue without AuNSs, we find $\Delta T_{ss}$ = 1.7 K (Fig. \ref{fig:2}(A)). For a tissue at an ambient temperature of 37$^{\circ}$ C this corresponds to moderate hyperthermia. In contrast, we measure extreme hyperthermia ($\Delta T> 5^{\circ}$ K) in the presence of $3.2\times10^8$ particles/ml and a laser power of 1.5 W: $\Delta T_{ss}=7.8$ K. Hence, there is a pronounced difference in the temperature achieved and, hence, of the effect of the PTTP, depending on whether or not AuNSs are present in the phantom.  

In a similar experiment (Fig. \ref{fig:2}(B)) we kept the AuNS concentration constant at 3.2$\times 10^8$ particles/ml and varied the laser power in the range from 0.5 W to 2 W. We find that the heating rate is ($4.9\pm0.7$) K/W. 

\subsection*{NIR laser-induced heating of cerebrum, cerebellum, and brain stem tissue}
\begin{figure}
\includegraphics[]{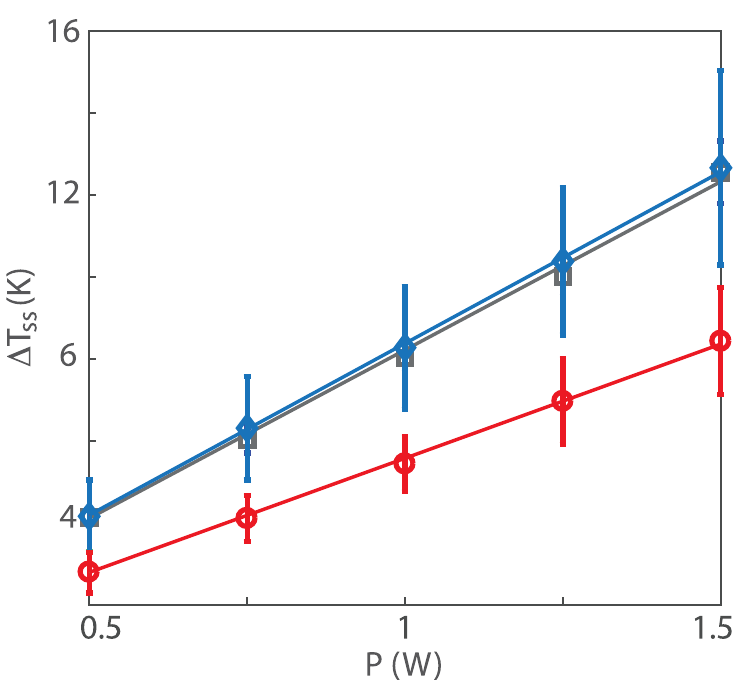}
\caption{\label{fig:3}\textbf{NIR laser-induced heating of brain tissue}. Steady-state temperature $\Delta T_{ss}$ versus laser power $P$ of non-homogenized brain stem (red circles), cerebrum (gray squares), and cerebellum (blue diamonds). The data is averaged ($N=5$) from different animals (except for cerebellum at 0.75 W, for brain stem at 0.5 W and at 0.75 W where $N=4$). The error bars denote one weighted standard deviations of the fitted values (Eq. \eqref{heatFit} and Eq. \eqref{coolFit}) of $\Delta T_{ss}$.}
\end{figure}

Paralleling the experiments in phantom tissue, we investigated the temperature increase of three different types of porcine brain tissue under laser irradiation. Due to the in-homogeneous distribution of grey matter and white matter in the different samples (Fig. \ref{fig:1}(B)) and in the cerebrum samples in particular, we consistently irradiate the samples from the same side to reduce the variance (cerebrum samples were irradiated from the cortex side). 
The thermal reaction of the brain tissue is represented by the steady-state temperature $\Delta T_{ss}$, which can be found by fitting the heating curve of the irradiated sample with Eq. \eqref{heatFit} and \eqref{coolFit}. The steady-state temperature $\Delta T_{ss}$ of the three tissue types, i.e., the cerebellum, cerebrum, and brain stem, is tested for laser powers from 0.5 W to 1.5 W, with an interval of 0.25 W and the results are shown in Fig. \ref{fig:3}, the error bars are weighted standard deviations of tissues from 5 different animals and the solid lines are linear fits to the data.
The observed temperature increases are - like the results obtained for the artificial tissue - linearly proportional to the laser power, $P$. From the slope the heating rate can be deducted, and for cerebrum, cerebellum and brain stem we obtain heating rates of ($8.2\pm0.2$) K/W, ($8.4\pm0.1$) K/W, and ($5.6\pm$0.1) K/W, respectively. Notably, the heating rate of brain stem is significantly ($p<<0.05$) lower than the other two types of brain tissues, hence, this tissue type heats significantly less when irradiated. Cerebrum and cerebellum are found to exhibit very similar heating rates ($p=0.14$). Furthermore, the heating rates of all types of tested brain tissues are larger than for the investigated phantom tissue, which is ($1.4\pm$0.1) K/W.

\subsection*{Plasmonic photothermal heating of brain tissue}
\begin{figure*}
\includegraphics[]{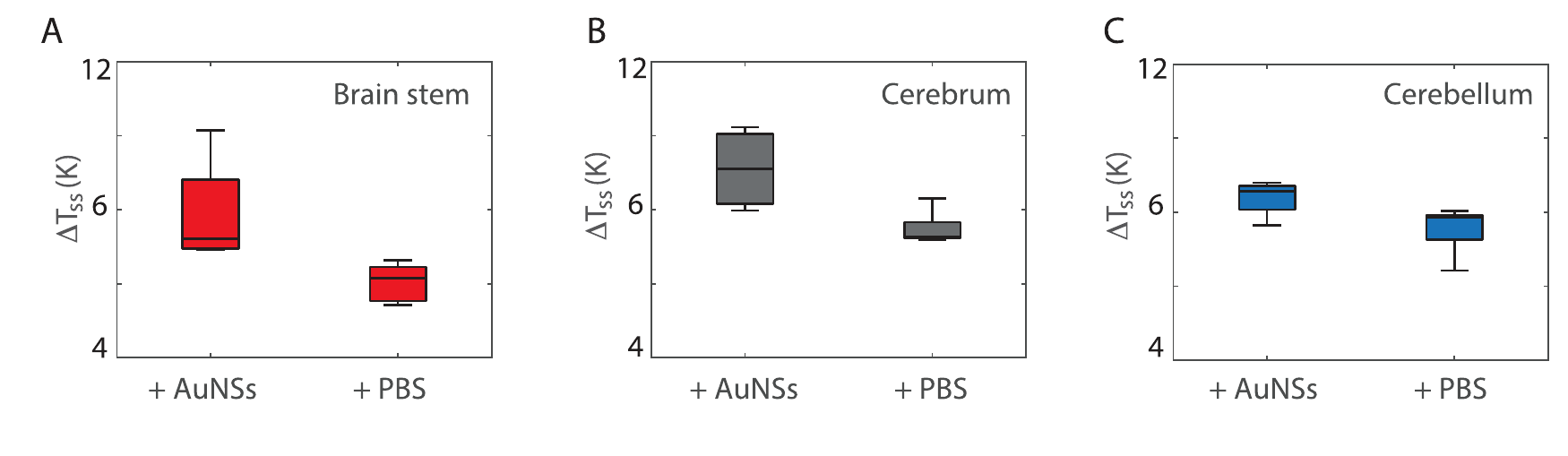}
\caption{\label{fig:4}\textbf{Plasmonic photothermal heating of brain tissue}. Box plot of the steady-state temperature, $\Delta T_{ss}$, during laser irradiation ($P=1$ W) of homogenized cerebral tissue injected with $3.3\times10^8$ particles/ml AuNSs (+AuNSs) compared to a control injected with saline solution (+PBS). $N=5$ (+PBS) and $N=4$ (+AuNS). (\textbf{A}) Brain stem ($p = 0.038$). (\textbf{B}) Cerebrum ($p = 0.021$). (\textbf{C})  Cerebellum ($p = 0.073$).
}
\end{figure*}

To investigate the heating of laser-irradiated brain tissues with plasmonic nanoparticles, we first injected AuNS into the tissue. However, it was challenging to distribute the injected particles and we did not measure any significant difference in $\Delta T_{ss}$ wit or without AuNSs. Probably, this was due to the very inhomogenous distribution of the AuNSs in the tissue. 
In order to obtain a more homogeneous distribution of nanoparticles in the samples, we instead homogenized the brain tissue using a centrifugal stirrer (blender) before transferring the tissue to the cuvettes. The results of these experiments are shown in Fig. \ref{fig:4}, which, for all three brain tissue types, displays the equilibrium temperature $\Delta T_{ss}$, with (+AuNSs) and without AuNSs (+PBS). Data point represents  4 or 5 independent experiments of each tissue type. Each sample is a mixture of tissue from two animals injected with either AuNS to a final concentration of $3.3\times10^8$ particles/ml or with the same volume of PBS. As expected, we found that the presence of plasmonic AuNS significantly augment $\Delta T_{ss}$ compared to the control ($p=0.038$, $p=0.021$ for brain stem and cerebrum, respectively) although the effect is slightly less pronounced for cerebellum ($p=0.073$). Still, however, for all tissue types the effect is small compared to the experiments performed in artificial tissue where we found $\Delta T>3$ K (Fig. \ref{fig:2}). One obvious difference between the brain tissue and the artificial tissue, i.e, agarose, is the transparency. While the first is very dense, the artificial tissue is just slightly opaque and, hence, lets the majority of the laser light pass through the sample. To investigate this further, we measured the intensity loss in the tissue. 

\subsection*{Absorbance spectroscopy in brain tissue}
\begin{figure}
\includegraphics[width=0.45\textwidth]{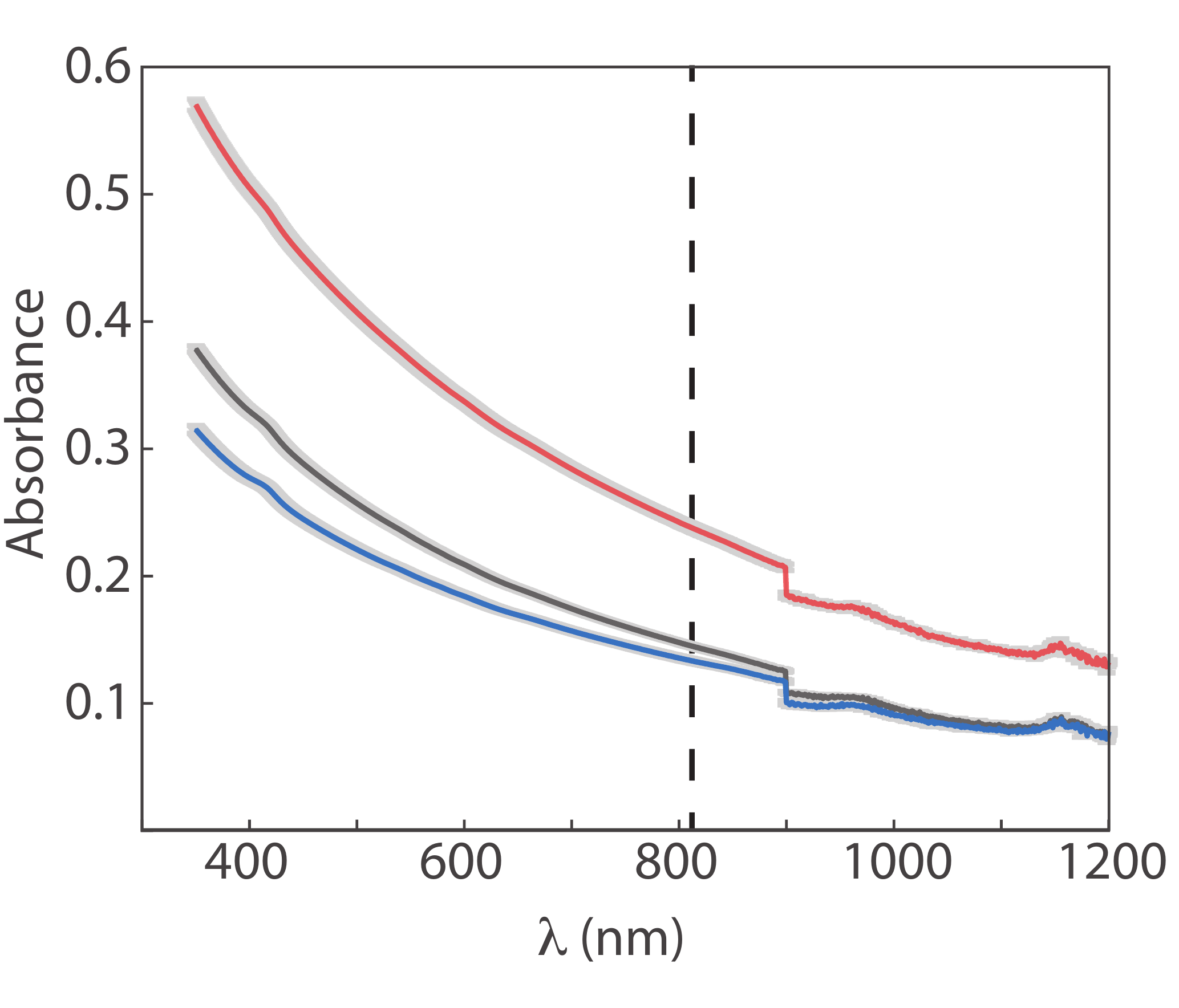}
\caption{\label{fig:5}\textbf{UV-Vis-NIR absorbance spectra of homogenized porcine brain}. Absorbance versus wavelength and error bars (light gray) show standard deviations ($N=3$). Brain stem (red), cerebrum (gray), and cerebellum (blue) show similar transmittance curve, but are vertically shifted with the brain stem having the highest absorbance, followed by the cerebrum and the cerebellum who show similar absorbance. The vertical black punctuated line signifies the wavelength (806 nm) of our laser and the jump at 900 nm is an artifact from the shift of detectors at this wavelength.}
\end{figure}

With photospectrometry in the ultraviolet to near-infrared range (UV-Vis-NIR) the absorbance of the homogenized porcine brain was evaluated. The absorbance is a dimensionless measure of the concentration dependent attenuation or the loss of laser intensity in the tissue. To ensure accuracy, the absorbance was kept below 1 by diluting the homogenized brain tissue in a saline solution (PBS). UV-Vis-NIR spectra for the three tissue types can be seen in Fig. \ref{fig:5}. Although the spectra have similar form, the magnitude of absorbance varies. The high absorbance of the brain stem is contradicted by the heating properties illustrated in Fig. \ref{fig:3}, where the brain stem consistently shows a lower increase in temperature when irradiated by a laser. Since the brain stem does not absorb laser light to as high a degree as the other examined brain parts, scattering is suspected to be the main contributor to the measured absorbance.
The curves in Fig. 5 show no signs of the absorption peaks outside of the biological window, which is in agreement with the conclusion that the absorption is dwarfed by scattering.

\section*{Discussion}
The effect of NIR laser radiation on biological tissue is a complex interplay between distinct phenomena; light-to-heat conversion, heat conductivity, and tissue reactions, e.g., photo-chemistry. The latter leads to denaturation - or destruction - of the tissue under treatment and depends on both laser (wavelength, power, beam profile) and on the tissue itself (optical coefficients and thermal conductivity).
As the thermal conductivities of the different cerebral tissues are approximately the same \cite{Murphy2016}, we can assume the temperature rise is dominated by the loss of intensity through the tissue. 

We found that the heating rate of brain stem (5.6 K/W) was significantly lower that the heating rates of the cerebrum (8.2 K/W) and cerebellum (8.4 K/W), which is suggestive of either lower absorption or higher scattering of brain stem tissue. This was further validated by measurements of the dimensionless absorbance, which describes the attenuation of the light in the tissue caused by both scattering and absorption, and this was higher for brain stem than the other cerebral tissue types investigated.
Our observation that scattering is high for brain stem tissue is in line with similar investigation of the optical properties of homogenized brain tissue \cite{Eisel2018}. Furthermore, these results are in correspondence with measured scattering coefficients of many different fatty tissues, see Ref. \cite{Jacques2013} for a thorough review on the subject. 
Hence, we measured significantly higher absorbance of brain stem tissue compared the other two tissue types over the entire spectrum (300 - 1200 nm).
Brain stem contains more white matter than cerebrum and cerebellum, which explains the more whitish colour of the brain stem (Fig. \ref{fig:1}(A-B)). Furthermore, it has been shown in different samples of white and grey brain matter (sliced) that white matter has substantially larger extinction coefficients than grey matter over the entire spectrum (300-1100 nm) \cite{Yaroslavsky2002}. It follows from this that our (806 nm) laser light penetrates deeper into the cerebrum and the cerebellum samples than the brain stem sample. Thus, more light can be absorbed in the cerebrum and cerebellum samples than in the brain stem, which explains why the brain stem heats less (Fig. \ref{fig:3} and Fig. \ref{fig:4}). 

The addition of plasmonic nanoparticles to the brain tissue increases the absorption coefficient of the whole sample. For the heating experiments, homogenizing the sample is beneficial in terms of comparing samples and reproducing results. In particular, for small heterogeneous samples from biopsies or resections.
However, homogenization of brain tissue is reported to cause an increase in absorption and decrease in scattering coefficients, which is attributed to the more even distribution of hemoglobin and to the destruction of membranes and other cellular structures when blending the tissue \cite{Eisel2018,Roggan1999}. 
Furthermore, our storage of tissue may have altered the optical properties of the samples, as there are indications that refrigerating for 24-48 hours can cause a decrease in absorption and an increase in scattering of homogenized porcine liver tissue. In contrast, slow freezing at -20$^{\circ}$ C was followed by a significant decrease in both absorption and scattering coefficients  \cite{Roggan1999}. In the current experiments, however, all samples were treated identically, hence, comparisons between the tissue types should be valid.  

Given this knowledge, PPTT is largely affected by scattering of the brain tissue. Thus, a possible implementation of the therapy in the brain must take this into regard and adjust the laser irradiation very locally, i.e., directly in the tumor. In contrast, there are indications that the absorption of NIR laser light in cerebral matter is lower than in tissues with high hemoglobin levels. For this reason, a highly localized laser and nanoparticle therapy is less likely to heat the surrounding healthy tissue. Hence, PPTT in the brain have the potential to provide a more localized treatment. Due to the risk of side-effects, e.g., hyperthermia and photochemistry, laser powers should be kept low to deposit a minimum of energy in healthy cerebral tissue. This, of course, is contrasted by the aim of maximizing hyperthermia in the tumor.
In conlucion, we show that hyperthermia of the brain matter around a tumor can not be neglected and a precise analysis of this local environment is needed to balance effect and side-effects prior to any potential therapies.  
%

\section*{Methods}
\subsection*{Artificial tissue and AuNSs mixture preparation} 
The AuNSs (peak absorbance at 800nm, mPEG coated, $2.9 \times 10^9$ particles/ml, in Milli-Q water, NanoXact NanoComposix) have silica cores ($\sim120$ nm diameter) and 15 nm gold shells. 
The artificial tissue was made of agarose powder (A9539-50G, SIGMA) dissolved in phosphate-buffered saline (PBS) to a concentration of 0.6\% (w/V) in a 100$^{\circ}$ C water bath for 2-5 minutes. Once the agarose powder was dissolved, it was removed from the water bath and left to cool until the agarose solution reached $\sim$50$^{\circ}$ C. Then the dissolved agarose was mixed with AuNSs solution in 4.5 ml polystyrene optical cuvettes (Kartell, 634-8531) with dimensions of (1 cm)$\times$(1 cm)$\times$(4.5 cm). The mixed solution was left to cool and solidify on the bench. Three different concentrations of AuNSs in agarose were prepared by mixing the AuNSs with the agarose solution: $3.2 \times 10^8$ particles/ml, $3.2 \times 10^7$ particles/ml, and $3.2 \times 10^6$ particles/ml. The concentrations are 10-1000 times less than the concentrations used for \textit{in-vivo} experiments of plasmonic photothermal therapy \cite{Jorgensen2016}.

\subsection*{Brain sample preparation} 
The brain tissue was dissected from newly butchered (according to Danish law) pigs and separated in 3 distinct parts: cerebrum, cerebellum, and brain stem (Fig. \ref{fig:1}(B)). The fresh brain tissue was kept for 3 days at the maximum before it was filled into the 4.5 ml cuvettes.  We used a syringe to create an under-pressure in the cuvette in order to ease the transfer. All samples had similar mass 4.1 g $\pm 0.1$ g (mean$\pm$ SD). Before irradiation, the samples were kept at room temperature for about 2 hours for calibration.

\subsection*{Laser irradiation}
Inspired by Ref. \cite{Cole2009} and Ref. \cite{Samadi2018} we sat up a laser irradiation and thermal imaging system based on the following components: An 806 nm laser (Modulight Inc., ML6600-0A1) that is transmitted through an optical cable connected with an illumination kit (Modulight Inc., MLAKIT EID655369) including a lens and a horizontally placed protecting shield. We confirmed the laser power using a standard power meter (P-link; Gentec, Sweden) and measured the beam width ($1/e^2$) to be $\sim1$ cm, which is the width of our cuvettes. We used an InSb IR camera (FLIR Systems SC4000, Boston, MA) with a spectral range of 3-5 $\mu$m, to measure the heating characteristics of the sample. The IR camera was mounted at an angle of approximately 90$^{\circ}$ to the laser such that the sample was irradiated completely while not interfering with the imaging. Each sample was illuminated at a constant power in the range from 0.5-2 W and the frame rate was 6/minute. For heating-cooling cycles, we irradiated the samples for 30 minutes and let them cool to room temperature for another 30 minutes, after the laser was turned off. During the experiments we monitored the variations in the ambient temperatures (see background in Fig. \ref{fig:1}(C)) and the standard deviations of the ambient temperatures were used to filter our data using Chauvenet's criteria for data filtering.

\subsection*{Heating of brain tissue with AuNSs}
In the experiments with blended brain, we compared the brain tissues from 6 different animals. The brain samples were prepared as described above, and then blended using a centrifugal stirrer (IKA\textregistered-WERKE RW16 Basic) and a Tissue Grinder with serrated pestle (Thomas\textregistered). The homogenized brain was transferred into cuvettes with volumes of about 3.4 cm$^3$. We washed the AuNSs solution by spinning and re-suspending in PBS. Our tissue samples were injected and mixed (vortex) with 207 $\mu$l of PBS with re-suspended AuNSs to a final concentration of $3.33 \times 10^8$ particles/ml. The control group was injected with an equal volume of PBS. 

\subsection*{Absorbance spectroscopy}
Spectrophotometry was performed using a Cary 5000 UV-Vis-NIR spectrophotometer (Agilent). Homogenized brain tissue from cerebrum, cerebellum, and brain stem were diluted in PBS (Thermo Scientific). Dilutions of 1:2000 (w/V) were used for spectrophotometry to ensure total absorbance below 1. Measurements were performed in quartz cuvettes to enable spectroscopy in the wavelength range from 350-1200 nm.

\subsection*{Data analysis}
We selected a region of interest in the thermal images to obtain the maximum temperature, using the ThermoVision software (FLIR Sys., Boston, MA) as shown in Fig. \ref{fig:1}(C). 
To analyze the extracted data, we fitted the expressions given in Eq. \eqref{heatFit} and Eq. \eqref{coolFit} to data. Following the train of thought from Ref. \cite{Roper2007} we obtained the expressions from an investigation of the energy balance of the system:
\begin{equation}\label{eq:5}
\sum_i m_iC_{p,i}\frac{dT}{dt}=\sum_j Q_j
\end{equation}
where the left side is the sum of products of masses, $m_i$, and the corresponding heat capacity, $C_{p,i}$, of the different components, $T$ is the temperature and $t$ is time. The right hand side is the sum of energy terms, $Q_j$. Eq. \eqref{eq:5} is valid when the time it takes for the system to reach thermal equilibrium within the cuvette is less than the time needed to obtain thermal equilibrium with the surroundings. The source term $Q_1$ is the heat dissipated by electron-phonon relaxation of plasmons on the nanoparticle surface at the laser wavelength, $\lambda$,      
\begin{equation}\label{eq:Q1}
Q_1=P(1-10^{-A_{\lambda}})\eta,
\end{equation}
where $\eta$ is the photothermal transduction efficiency, i.e., the efficiency to convert the incident absorbance, $A_{\lambda}$, of laser light to heat. $A_{\lambda}$ is given by Beer-Lambert's law and is often referred to as the optical density. $Q_0$ is the heat dissipated in the cuvette and the media containing the nanoparticles. The terms $Q_1$ and $Q_0$ add heat to the system and are counteracted by energy terms describing exchanges with the surroundings. The first term is energy conducted to air, sample holder etc.:
\begin{equation}
Q_{cond}\propto \Delta T, 
\end{equation}
where $\Delta T=T-T_{amb}$ and $T_{amb}$ is the ambient temperature. Furthermore, there is thermal radiation, $Q_{rad}$, given by Stefan-Boltzmann's law:   
\begin{equation}\label{eq:Qrad}
Q_{rad}\propto T^4-T_{amb}^4
\end{equation}
For large $T$, i.e., $T>T_{amb}$, $Q_{rad}/\Delta T$ varies little respect to $\Delta T$. For $\Delta T$ less than $10^{\circ}$ K we can approximate $Q_{rad}/\Delta T$ with a constant value as the variation is less than 5\% \cite{Roper2007}. As a result, 
\begin{equation}
Q_{ext}=Q_{cond}+Q_{rad}\propto\Delta T
\end{equation}
and, thus, is written as:
\begin{equation}\label{eq:Qext}
Q_{ext}=hA(T-T_{amb}),
\end{equation}
where $h$ is a heat-transfer coefficient and $A$ is the surface area for radiative heat transfer. Thus, eq. \eqref{eq:5} is simplified to:
\begin{equation}\label{eq:sum}
\sum_i m_i C_{p,i}\frac{dT}{dt}=Q_1+Q_0-Q_{ext}
\end{equation}

We define the system's characteristic time constant to be
\begin{equation}\label{eq:tau}
\tau:=\frac{\sum_i m_i C_{p,i}}{hA}
\end{equation}
To extract $\tau$ we focus on thermal equilibrium with the surroundings via conductive and radiative heat transfer, i.e. after laser is turned off. In this case $Q_1=Q_0=0$ and eq. \eqref{eq:sum} reduces to
\begin{equation}\label{eq:sumc}
\sum_i m_i C_{p,i}\frac{dT}{dt}=-Q_{ext}
\end{equation}
We define the dimensionless driving force to be
\begin{equation}
\theta:=\frac{T_{amb}-T}{T_{amb}-T_{ss}},
\end{equation}
where $T_{ss}$ is the maximum temperature reached or the steady-state temperature. Then, we substitute $\tau$ and $\theta$ into eq. \eqref{eq:sumc} and get 
\begin{equation}
\frac{d\theta}{dt}=-\frac{\theta}{\tau}
\end{equation}
and integrate
\begin{equation}
\int\frac{1}{\theta}d\theta=-\frac{1}{\tau}\int dt
\end{equation}
Using the initial condition that $\theta=1$ when $t=0$, i.e., when irradiation ceases, we get
\begin{equation}\label{tau2}
\log\theta=-\frac{t}{\tau}
\end{equation} 
such that
\begin{equation}
\theta=e^{t/\tau}
\end{equation}
By substituting $\theta$ we find the expression given in Eq. \eqref{coolFit}

When the laser is irradiating the system, heat is added ($Q_0+Q_1>0$) and we use the initial condition that $\theta=0$ when $t=0$ we find
\begin{equation}
\theta=1-e^{t/\tau}
\end{equation}
By substituting $\theta$ we get Eq. \eqref{heatFit}.

\begin{acknowledgments}
The authors acknowledge financial support from the Danish National Research Councils (DNRF116). YH acknowledges financial support from the Chinese Scholarship Council. The authors thank Harald Hansen's abattoir for providing the fresh porcine brain, Martin Roursgaard, Institute of Public Health, University of Copenhagen, for kindly providing the tissue blender, and Kamilla Nørregard for fruitful discussions.\\
\end{acknowledgments}


\section*{References}
\providecommand{\latin}[1]{#1}
\providecommand*\mcitethebibliography{\thebibliography}
\csname @ifundefined\endcsname{endmcitethebibliography}
  {\let\endmcitethebibliography\endthebibliography}{}

\end{document}